\documentclass[a4paper, 10pt, conference]{IEEEtran}
\usepackage{const4biclasso}
\usepackage{amsmath,graphicx}
\usepackage{scrextend}
\usepackage{makecell}
\usepackage{epstopdf}
\usepackage[normalem]{ulem}
\begin{document}

\title{Simultaneous Signal Subspace Rank and \\Model Selection with an Application to \\Single-snapshot Source Localization}

\author{\IEEEauthorblockN{Muhammad Naveed Tabassum and Esa Ollila}
\IEEEauthorblockA{Aalto University, Dept. of Signal Processing and Acoustics, P.O. Box 15400, FI-00076 Aalto, Finland} 
Email: \{muhammad.tabassum, esa.ollila\}@aalto.fi }

\maketitle

\begin{abstract}
This paper proposes a novel method for model selection in linear regression by utilizing the solution path of $\ell_1$ regularized least-squares (LS) approach (i.e., Lasso). This method applies the complex-valued least angle regression and shrinkage (c-LARS) algorithm coupled with a generalized information criterion (GIC) and referred to as the c-LARS-GIC method. c-LARS-GIC is a two-stage procedure, where firstly precise values of the regularization parameter, called knots, at which a new predictor variable enters (or leaves) the active set are computed in the Lasso solution path. Active sets provide a nested sequence of regression models and GIC then selects the best model. The sparsity order of the chosen model serves as an estimate of the model order and the LS fit based only on the active set of the model provides an estimate of the regression parameter vector. We then consider a source localization problem,  where the aim is to detect the number of impinging source waveforms at a sensor array as well to estimate their direction-of-arrivals (DoA-s) using only a single-snapshot  measurement.  We illustrate via simulations that, after formulating the problem as a grid-based sparse signal reconstruction problem, the proposed c-LARS-GIC method detects the number of sources with high probability while at the same time it provides accurate estimates of source locations. 
\end{abstract}


\section{Introduction}

We assume that a measurement vector $\y  \in\C^{n}$  is generated via linear model 
\beq\label{eq:linear} 
\y = \X \beb + \vepsb,
\eeq
where $\X=\bmat \x_1 & \cdots & \x_p \emat \in\C^{n\times p}$ is the known basis matrix (or matrix of predictors), $\beb  \in\C^{p}$ is the  unknown signal vector (or regression coefficient vector)  and $\vepsb \in\C^{n}$ is the (unobserved) random noise vector.  We assume that the model is  underdetermined, i.e., $\pdim > \ndim$, and the signal vector  $\beb$ is {\em sparse} (i.e., having only a few non-zero elements)  with sparsity order  $k^* = \|\beb\|_0  \ll n$,  where $\| \cdot \|_0$ denotes the $\ell_0$-(pseudo)norm, $\| \beb \|_0 = | \{i  \, : \,  \beta_i\neq 0\}| $, i.e.,  $\| \beb \|_0$ is equal to number of non-zero elements of $\beb$.  Such a model arises  in compressed sensing \cite{candes2006robust} and in sparse linear regression \cite{hastie2015stat}. For ease of exposition, we consider centered linear model,  i.e., the intercept is equal to zero.

In this paper, we consider solving the following tasks
\begin{itemize}
\item {\it Detection task}, where the problem is to find the true model (sparsity) order $k^*$,  or in other words, to detect the rank of the signal subspace. 
\item {\it Sparse signal reconstruction (SSR) task}, where the problem is to estimate the unknown $k^*$-sparse signal $\beb$.  
\end{itemize} 
So far most research has focused on SSR task and much less attention has been paid to solve the detection task. Some existing  detection approaches in the literature  can be found e.g., in  \cite{romer2014sparsity,austin2010relation}. Typically, most SSR (e.g.,  greedy pursuit methods  \cite{tropp2007signal})  presume that $k^*$ is known or an estimate of it is available.  
In practice, the underlying sparsity order $k^*$ is typically unknown and can even vary with time, e.g., as new measurements (or snapshots) become available.  Thus the performance of greedy pursuit methods depends heavily on the accuracy of the method used for solving the detection task.  Some other approaches,  such as sparse linear regression methods \cite{hastie2015stat}, perform detection and SSR task simultaneously. For example, the Least absolute shrinkage and selection operator (Lasso)  \cite{lasso:1996} uses a data-dependent penalty parameter  $\lam$ commonly chosen by cross-validation to compute an estimate $\hat \beb$ of sparse vector $\beb$. An estimate of  the sparsity order $k^*$ is then  obtained as  $\hat k^* = \| \hat \beb\|_0 $.

The Lasso estimator of $\beb$  solves the following penalized (regularized) residual sum of squares (RSS) optimization problem,
\beq\label{eq:penEN} 
 \bebh(\lam) = \underset{\beb \in \C^\pdim}{\arg \min}  
\ \frac  1 2  \|\y - \X \beb \|_2^2   + \lam \|\beb\|_1 
\eeq 
where $\lam \geq 0$ is the penalty parameter that controls the sparsity order of the solution. The {\sl least angle regression and shrinkage (LARS)} algorithm  \cite{efron2004LAR} is a novel method for computing the  Lasso solution (regularization) path as $\lam$ varies. The LARS method finds the sequence of values $\lam_k$, referred to as {\sl knots}, at which a new predictor variable enters (or leaves) the active set of non-zero coefficients. Additionally, it provides the respective solutions $\bebh(\lam_k)$ for $k=0,1,\ldots, K = \min(n-1,p)$. Herein, we use our complex-valued extension of LARS (see  \cite{tabassum2018sequential,saenlars}) to find the knots and the associated active sets of non-zero coefficients in the Lasso solution path \cite{efron2004LAR, tabassum2017pathwise, tabassum2018sequential}. Active sets then provide a nested sequence of regression models. At the second stage, we then apply an information criterion (IC) to choose the best model among these candidate models.  This approach is referred to as  {\sl c-LARS-GIC} method in the sequel. 

There are several information criteria (e.g.,  Akaike or Bayesian or their extensions) proposed in the literature (see \cite{stoica2004model} for a review) that can be utilized at the second stage.  Another alternative would be to use  sequential generalized likelihood ratio tests \cite{koivunen2014model}, the covariance test \cite{lockhart2014} or spacing test \cite{spacingsTest2016}, for example. However, we choose to use information criteria due to their simplicity and ease of implementation. Furthermore, tests developed in \cite{lockhart2014,spacingsTest2016} work only for real-valued data.

In this paper, we use  {\sl generalized information criterion (GIC)} formulation \cite{nishii1984} that contains both the Bayesian information criterion (BIC) \cite{schwarz:1978},  and its modifications developed  in \cite{Wang2009,FanTang2013}  as special cases. The sparsity order of the chosen model serves as an estimate of the model order, and the least-squares (LS) fit based on the active set of the chosen model provides estimates of the non-zero signal coefficients. Thus, {\sl c-LARS-GIC} approach solves both the detection and estimation task simultaneously. 

This paper is organized as follows. Section~\ref{sec:LARS-BIC} develops the c-LARS-GIC method. The grid-based SSR framework for single-snapshot  source localization problem using a sensor array, referred to as compressed beamforming (CBF),  is described in Section~\ref{sec:cbf}. The results for a set of simulation studies then illustrate the effectiveness of the c-LARS-GIC approach in CBF application.  Section~\ref{sec:concl} discusses the key outcomes and concludes the paper. 

{\it Notations}: 
Lowercase boldface letters are used for vectors and uppercase for matrices.
The $\ell_2$ and $\ell_1$ norms of the vector $\a \in \C^\pdim$ are defined as $\| \a \|_2 = \sqrt{\a^\hop \a}$ and  $\| \a\|_1 = \sum_{j=1}^\pdim |a_i |$, where $| a | = \sqrt{a^* a} = \sqrt{a_R^2 + a_I^2} $ denotes the modulus of a complex number $a=a_R + \im a_I$  and $(\cdot)^\hop=[(\cdot)^*]^\top$ denotes the Hermitian  (complex conjugate) transpose. The support of $\a$ is the index set of its non-zero elements, i.e., $\setA = \supp(\a)= \{ j \in \{ 1, \ldots, \pdim\} : a_j \neq 0 \} $. The $\ell_0$-(pseudo)norm of  $\a$ is defined as $\| \a\|_0 = | \supp(\a)|$, which is equal to the total number of non-zero elements in it.  
If $\beb \in \C^\pdim$, $\X \in \mathbb{C}^{\ndim \times \pdim}$ and $\setA$ is an index set $\setA \subset \{1, \ldots, \ndim\}$ of cardinality $| \setA | = k$, then $\beb_\setA \in \C^k$ (resp.  $\X_{\setA}$) denotes the sub-vector (resp. $\ndim \times \kdim$ sub-matrix)   
restricted to components of $\beb$ (resp. columns of $\X$) indexed by the set $\setA$. The $\X^+$ is a Moore-Penrose pseudoinverse of $\X$.
We use  $\mathsf{I}(\cdot)$  to denote the indicator function and $\mathrm{ave} = \frac{1}{M}\sum_{m=1}^M (\cdot)$ for $M$ numbers.

\section{The c-LARS-GIC Method} \label{sec:LARS-BIC}

\subsection{The Lasso Knots and Active Sets}

Recall that the signal vector $\beb \in \C^\pdim$  is sparse, so only few elements are non-zero. The support $\setA^*$ of $\beb$ is the index set of its non-zero elements, i.e., $\setA^* = \supp(\beb) $ and the cardinality of this set, $k^* = \| \beb \|_0 <\ndim < \pdim$, is the true sparsity level or the model order. The main aim in detection task is to find the true order $k^*$ whereas the main interest in SSR task is on accurate variable selection, i.e., on identifying the true active set $\setA^*$. Naturally, if $\setA^*$ would be known, then one can simply compute the LS solution when regressing $\y$ on  $\X_{\setA^*}$, where $\X_{\setA^*}$ denotes the  $\ndim \times k^*$ (sub)matrix of  $\X$ restricted to the columns of  the active set $\setA^*$. 

Recall that Lasso estimate $\hat \beb(\lam) \in \C^\pdim$ is sparse having {\it at most $n$ non-zero elements}. The sparsity level of $\hat \beb(\lam)$ depends on the penalty parameter $\lam$. We assume that the columns of $\X$ are centered, i.e., $\| \x_j \|^2 = 1$ holds.  In Lasso estimation framework this can be done without loss of generality since it only implies that the found Lasso solution (based on normalized basis vectors)  is rescaled back to the original scale of $\x_j$-s; see \cite{hastie2015stat} for more elaborate discussion of this feature of the Lasso.

Let $\lam_0$ denotes the smallest value of $\lam$ such that all coefficients are zero, i.e., $\hat \beb(\lam_0)=\bo{0}_\pdim$.  It is easy to see that  \cite{hastie2015stat}
\beq \label{eq:lam0}
\lam_0= \max_{j \in \{1, \ldots, p\}} | \langle \x_j, \y \rangle | .
\eeq
Let $\setA(\lam) = \supp\{\hat \beb(\lam)\}$ denotes the {\paino active set} at the regularization parameter value $\lam < \lam_0$. 
The {\paino knots}  $\lam_0> \lam_1> \lam_2 > \cdots > \lam_K$ are defined as the smallest values of the regularization parameter after which there is a change in the set of active predictors, i.e.,  the order of sparsity changes.   The active set at a knot $\lam_k$ is denoted by $\setA_{k} = \setA(\lam_k) =  \supp\{\hat \beb(\lam_k)\}$.   
The active set  $\setA_1$ thus contains a single index as $\setA_1 = \{ j_1 \}$, where $j_1$ is predictor that becomes active first, 
i.e., 
\[
j_1 = \underset{j \in \{1, \ldots, p\}}{\arg \max}  | \langle \x_j , \y \rangle | . 
\]
 By definition of the knots, one has that 
 \begin{align*}
 \setA_k &= \supp \big\{\hat \beb(\lam) \big\} \;\,\;\quad  \forall \, \lam_{k-1} < \lam  \leq \lam_k \\ 
\setA_{k} &\neq \setA_{k+1}, \qquad\qquad\: \forall \, k=0,1,\ldots,K
 \end{align*} 
and $\setA_0 = \supp \big\{\hat \beb(\lam) \big\}  =\{ \emptyset \}$ for all $\lam \geq \lam_0$.

Let $\setA=(i_1,\ldots,i_k) \subset \{1, \ldots, \ndim\}$ denotes an index set of  variables (basis vectors $\x_j$-s) that are included in the model and suppose that  its sparsity order is $k = | \setA | < \ndim$. Then $\setA^\complement = \{1, \ldots, p\} \setminus \setA$ is the index set of signal coefficients that are zero. Then, ideally, we  would like to consider all hypotheses of the form
\beq \label{eq:hypo}
H_{\setA}  :  \beb_{ \setA^\complement}  = \bo 0, \qquad  | \setA | = k \: \implies \:  | \setA^\complement | = \pdim - k
\eeq 
and choose the best model among all possible models $\{H_\setA\}$ as the one that appears most plausible based on the data. The posed hypotheses testing problem is obviously computationally infeasible even for small $\ndim$.  Indeed, observe that there are $\ndim \choose k$  index sets $\setA$ of size $|\setA |=k$, where $k=1,\ldots,\ndim$. Thus it is not possible to go through all possible models since the total number of all models is  $O(\ndim^\ndim$).

Finding the Lasso knots $\lam_0,\lam_1,\ldots, \lam_K$ and  the corresponding active sets $\setA_k$ for $k =0,1, \ldots, K$, via the c-LARS-WLasso algorithm of \cite{tabassum2018sequential,saenlars} with unit weights allows us to form a nested set of hypotheses, 
\beq \label{eq:nested_hyp}
\begin{aligned}
&\Hp_0  \subset \Hp_1 \subset \cdots \subset \Hp_K \\ 
&\Hp_k  : \beb_{\setA_k^\complement}  = \bo 0,  \qquad k =0,1, \ldots, K. 
\end{aligned} 
\eeq 
This reduces the number of tested hypotheses from $\ndim^\ndim$ to $(K + 1)  \leq \ndim \leq \pdim$.

\subsection{Model Selection using Information Criteria}

Information criteria are a common way of choosing among models while balancing the competing goals of goodness-of-fit and parsimony (simpler model).
The  Generalized information criterion (GIC) \cite{nishii1984} for hypothesis $H_\setA$ is  
\beq \label{eq:BICreg}
 \mathrm{GIC}_{\gamma}(\setA) =   n \ln \hat \sigma^2(\setA) + k \, c_{n,\gam} 
\eeq 
where $k=| \setA |$ is the sparsity (model) order of hypotheses $H_\setA$, and 
\[
\hat \sigma^2(\setA) = \frac{1}{\ndim}  \big\| \y - \X_\setA \hat{ \bo s}  \big \|^2_2 , \quad \hat{ \bo s}  = \X_{\setA}^+ \y
\]
are the ML-estimators of error scale $\sigma^2>0$ and $\beb_{\setA^*}$ under $H_\setA$ in \eqref{eq:hypo} and assuming that $\vepsb \sim \mathcal N_\ndim(\bo 0, \sigma^2 \bo I)$. 
Furthermore,   $c_{n,\gam}$  in \eqref{eq:BICreg} is a positive {\it sequence} depending on $n$.    
Subscript $\gam$ obtains values $\gam \in \{0,1,2\}$ and is used to indicate the different options of sequence $c_{n,\gam}$ suggested in the literature:
\begin{itemize}
\item $\mathrm{GIC}_0$ = BIC  \cite{schwarz:1978}  uses $c_{n,0} = \ln \ndim$. 
\item $\mathrm{GIC}_1$ uses $c_{n,1} = \ln \ndim \cdot \ln(\ln \pdim)$ as  suggested in  \cite{Wang2009}.
\item $\mathrm{GIC}_2$ uses $c_{n,2} = \ln \pdim \cdot \ln(\ln \ndim)$ as suggested in \cite{FanTang2013}. 
\end{itemize}
If we assume that the set of hypotheses $ \{H_\setA\}$ 
includes the true model $H_{\setA^*}$ that generated the data, then the GIC estimates used above have proven to be consistent \cite{nishii1984,FanTang2013,Wang2009} under some technical conditions, that is,  
$\mathbb{P}(\mbox{``correctly  choosing $H_{\setA^*}$''}) \to 1$ as $\ndim \to \infty$. 
Finally, we point out that the ML estimate  $\hat \sigma^2=\hat \sigma^2(\setA)$ of $\sigma^2$ under the hypothesis $H_\setA$ is biased. Therefore, we instead use the bias corrected (i.e., unbiased) estimate, given as:
\beq \label{eq:sigma2_u}
\begin{aligned}
 \hat \sigma^2_u(\setA) &= \frac{1}{\ndim - k}  \big\| \y - \X_\setA \hat{\bo \s} \big \|^2_2, \\
 k&= | \setA|, \qquad  \hat{ \bo s}  = \X_{\setA}^+ \y. 
\end{aligned}
\end{equation}
Note that we will use $\hat \sigma^2_u(\setA_k)$ in place of $\hat \sigma^2(\setA)$ in \eqref{eq:BICreg}, for nested hypothesis set $\Hp_k$ of \eqref{eq:nested_hyp}, where $k =0,1, \ldots, K$.

\subsection{The c-LARS-GIC Method}

We are now ready to state the c-LARS-GIC procedure for testing the nested set of hypotheses in \eqref{eq:nested_hyp} for choosing the correct model and the sparsity order.  The method is described in Algorithm~\ref{algo:LARS-BIC}. 

\setlength{\textfloatsep}{2pt}
\begin{algorithm}[!h]
	\SetAlgoHangIndent{0pt}
	\DontPrintSemicolon
	\caption{The c-LARS-GIC method.}\label{algo:LARS-BIC}
	\SetKwInOut{Input}{input}\SetKwInOut{Output}{output}\SetKwInOut{Init}{initialize } 
	\Input{$\quad \y\in\C^{n}$, $\X\in\C^{n\times p}$ and $\gam \in \{0,1,2\}$.}
		\BlankLine

	\Output{\quad $(\hat k^*, \hat \setA^*, \bebh)$.} 
	\BlankLine
	
	\Init{$\quad \bebh = \bo 0_{\pdim}$ and $K = \min(n-1,p)$.}
	\BlankLine
	 Compute the knots  $\lam_k$ and the corresponding active sets $\setA_k$ using the c-LARS-WLasso algorithm with unit weights, where $k=0,1,\ldots, K$; see  \cite{tabassum2018sequential} and \cite{saenlars}. 
	\BlankLine

Find the sparsity level  $k^*$ using \eqref{eq:BICreg} as: 
\[
\hat k^* =  \underset{  k \in \{0,1,\ldots, K\} }{ \arg \min }  \  \{ \mathrm{GIC}_\gam(\lam_k) = \,  n \ln \hat \sigma^2_u(\setA_k)  \, + k \, c_{n,\gam}  \, \},
\] 
where $\hat \sigma^2_u(\setA_k)$ is computed using \eqref{eq:sigma2_u}. 
Thus, an estimate of the active set is 
$
\hat \setA^*  = \setA_{\hat k^*}. $
	\BlankLine

Set $\bebh_{\hat \setA^*} = \X_{\! \hat \setA^*}^+ \y$ to get $\hat k^*$-sparse  estimate $\bebh \in \C^{\pdim}$. 
\end{algorithm}

It is important to note that the nested set of hypotheses \eqref{eq:nested_hyp} are not pre-fixed as usually but are based on the data. Naturally, the success of the c-LARS-GIC method depends on the ability of the Lasso to have the correct model $\setA^*$ among the active sets $\{\setA_k\}_{k=1}^K$  which are determined at the knots in the Lasso path. Nevertheless, when the true $\setA^*$ is included in the set of models, then c-LARS-GIC has a high probability of choosing the correct model due to consistency property of the used information criterion. This feature is indeed verified in our simulation studies in Section~\ref{sec:simul}.

\subsection{Conventional Approach}\label{sec:comp}

We compare the  c-LARS-GIC method  to conventional approach where Lasso estimates $\{ \bebh(\lam_l)\}_{l=0}^L$ are found on a  grid of $\lambda$ values, e.g., by cyclic coordinate descent algorithm \cite{hastie2015stat},
\[
\{\lam_0,\ldots,\lam_L\},  \quad \lam_0 > \lam_1> \ldots > \lam_{L},
\] 
where the sequence  $ \{ \lam_l  \}$ is monotonically decreasing 
from $\lam_0$ to $\lam_L  \approx 0$ on a log-scale. Note that $\bebh(\lam_0)=\bo{0}_p$, where  $\lam_0$ is  given in \eqref{eq:lam0}. 
 By default, we use  $\lam_L = \epsilon \lambda_0$, so $\lam_{j} = \epsilon^{j/L} \lam_0= \epsilon^{1/L} \lam_{j-1}$
 with  $\epsilon = 10^{-3}$ which is the default value used by Lasso routine of \texttt{MATLAB}\textsuperscript{\textcopyright}. We consider a dense grid of ($ L=100$) values of $\lam$. 

The conventional approach then uses the information criterion  to choose the best solution among these candidate solutions as follows. 
First, we compute the scale estimate of each solution as:
\[
\hat \sigma^2_u(\lam_l)= \frac{1}{\ndim -\|  \hat \beb(\lambda_l) \|_0  }  \big\| \y - \X \bebh(\lam_l) \|^2_2, \quad l=0,1,\ldots,L.
\]
Then we compute the GIC optimum (for all values of $\gam$)
\[
\ell = \underset{l \in \{0,1,\ldots,L\}}{\arg \min}  \{ \mathrm{GIC}_\gam(\lam_l) = \,  n \ln \hat \sigma^2_u(\lam_l)  \, +  \|  \hat \beb(\lambda_l) \|_0 \, c_{n,\gam} \}
\]
and choose $\bebh = \bebh(\lambda_{\ell})$ as GIC based Lasso solution which gives $\hat k^* = \|  \hat \beb(\lambda_\ell) \|_0$.

\section{Compressed Beamforming Application}\label{sec:cbf}

\subsection{Single-snapshot Source Localization Problem}

Consider a sensor array processing application in which a uniform linear array (ULA) of $\ndim$ sensors is used for estimating the direction-of-arrivals (DoA-s)
of the sources with respect to the array axis.  The array response (steering vector) of ULA for a source from DoA (in radians) $\theta \in [-\pi/2, \pi/2) $  
is given by 
\[
\a\bigl(\theta\bigr)= \cfrac{1}{\sqrt{n}}\: \bigl(1,e^{ \im \pi\sin\theta},\ldots,e^{ \im \pi(n-1)\sin\theta}\bigr)^\top,  
\]
where we assume half a wavelength inter-element spacing between sensors. 
 We consider the case that $k^* < \ndim$ sources from distinct DoAs $\theta_1,\ldots,\theta_{k^*}$ 
 arrive at a sensor at some time instant $t$. 
A single snapshot obtained by ULA can then be modeled as 
\beq \label{eq:snapshot} 
\y = \bo A(\bom \theta) \bo s + \vepsb
\eeq 
where $\bo s \in \mathbb{C}^{k^*}$ contains the source waveforms at time instant $t$, $\bom \theta=(\theta_1,\ldots,\theta_{k^*})^\top$ collects the DoAs,  
and $\bo A =  \big(\a(\theta_1) \cdots \a(\theta_{k^*}) \big) \in \C^{\ndim \times k^*}$ is the array steering matrix and $\vepsb \in \C^\ndim$ is the complex random noise.

Consider an angular grid of size $\pdim$ (commonly $\pdim \gg \ndim$) 
of look directions of interest in terms of the broadside angles, 
\[
[\theta] = \{ \theta_{[j]} \in [-\pi/2, \pi/2) \ :  \   \theta_{[1]} <  \cdots < \theta_{[\pdim]} \}. 
\]
Let the $j^{th}$ column of  the measurement matrix $\X$  in the model \eqref{eq:linear} be the array response for look direction $\theta_{[j]}$, so $\x_j = \a(\theta_{[j]})$.   Then if the true source DoAs are contained in the angular grid, i.e.,  $ \theta_k \in [\theta]$ for $k=1,\ldots,k^*$, then the snapshot $\y$ in  \eqref{eq:snapshot} can be equivalently modeled as in \eqref{eq:linear} where $\beb$ is exactly $k^*$-sparse (i.e., $\| \beb \|_0=k^*$) and  non-zero elements of $\beb$  maintain  the source waveforms $\s$, i.e., $\beb_{\setA^*} = \bo s$, where $\setA^* = \supp(\beb)$. 
Thus, identifying the true DoA-s is equivalent to identifying the non-zero elements of $\beb$. 
This principle is known as {\sl compressed beamforming (CBF)}.  Hence, based on a single snapshot  only, c-LARS-GIC can be utilized for detecting the number of sources $k^*$ and the corresponding source parameters, such as DoAs as locations in the grid $[\theta]$ specified by $\setA^*$ and the source powers by $|s_k|= \big|[\beb_{\setA^*}]_k \big|$ for $k=1,\ldots, k^*$. 

\subsection{Simulation Studies and Results}\label{sec:simul}

In CBF, we apply the developed methods for estimating the true source parameters, e.g., number of sources (i.e., sparsity order) $k^*$ and their DoAs and powers. 
In this paper, we have grid $[\theta]$ for  a grid-spacing $\Del \theta = 2^{\circ}$ which gives $p=90$ possible DoAs in $\theta  \in [-90^{\circ},90^{\circ})$. 
The source waveforms $\s$ are generated as $s_k =  |s_k | \cdot e^{ \im  \mathrm{Arg}(s_k)}$ for $k=1,\ldots,k^*$, where source powers $|s_k| \in (0,1]$ are fixed but the source phases are randomly generated for each of $M$ Monte-Carlo (MC) trials as $\mathrm{Arg}(s_k) \sim \mathrm{Unif}(0,2 \pi)$.  
The error terms $\varepsilon_i$ are independent and identically distributed and generated from $\C \mathcal N(0,\sigma^2)$ distribution, where $\sigma^2$ is chosen to yield desired signal-to-noise ratio (SNR) in decibels (dB),  $\mathrm{SNR}$(dB) $  = 10\log_{10} (\sigma_s^2/\sigma^2)$, 
where $\sigma_s^2 = \frac{1}{\kdim^*} \bigl\{|s_1|^2 +|s_2|^2+\dots+|s_{\kdim^*}|^2\bigr\}$ denotes the average source power.

The performance measures are the (empirical) {\sl probability of detection},  
$
\mathrm{PD} =  \mathrm{ave} \{ \mathsf{I} ( k^*= \hat k^*)\}
$
and the empirical probability of exact recovery, 
$
\mathrm{PER}  =  \mathrm{ave} \{   \mathsf{I} \big ( \setA^* = \hat \setA^*\big) \}. 
$
We also report the mean squared error, 
$
\mathrm{MSE} =  \mathrm{ave}  \{  \big\| \bo s   - \bebh_{\hat \setA^*} \big\|^2_2\}
$
to illustrate the performance in estimation of the source powers. 
Above the average is over $M=1000$  MC trials and $\hat k^*$, $\hat \setA^*$ and $\bebh_{\hat \setA^*}$ denote the estimates of $k^*$,  $\setA^*$ and $\beb_{\setA^*}$  obtained by running the  Algorithm~\ref{algo:LARS-BIC} for  a given MC data set.

\begin{figure}[!t]
	\centering
	\subfloat{\includegraphics[width=\columnwidth]{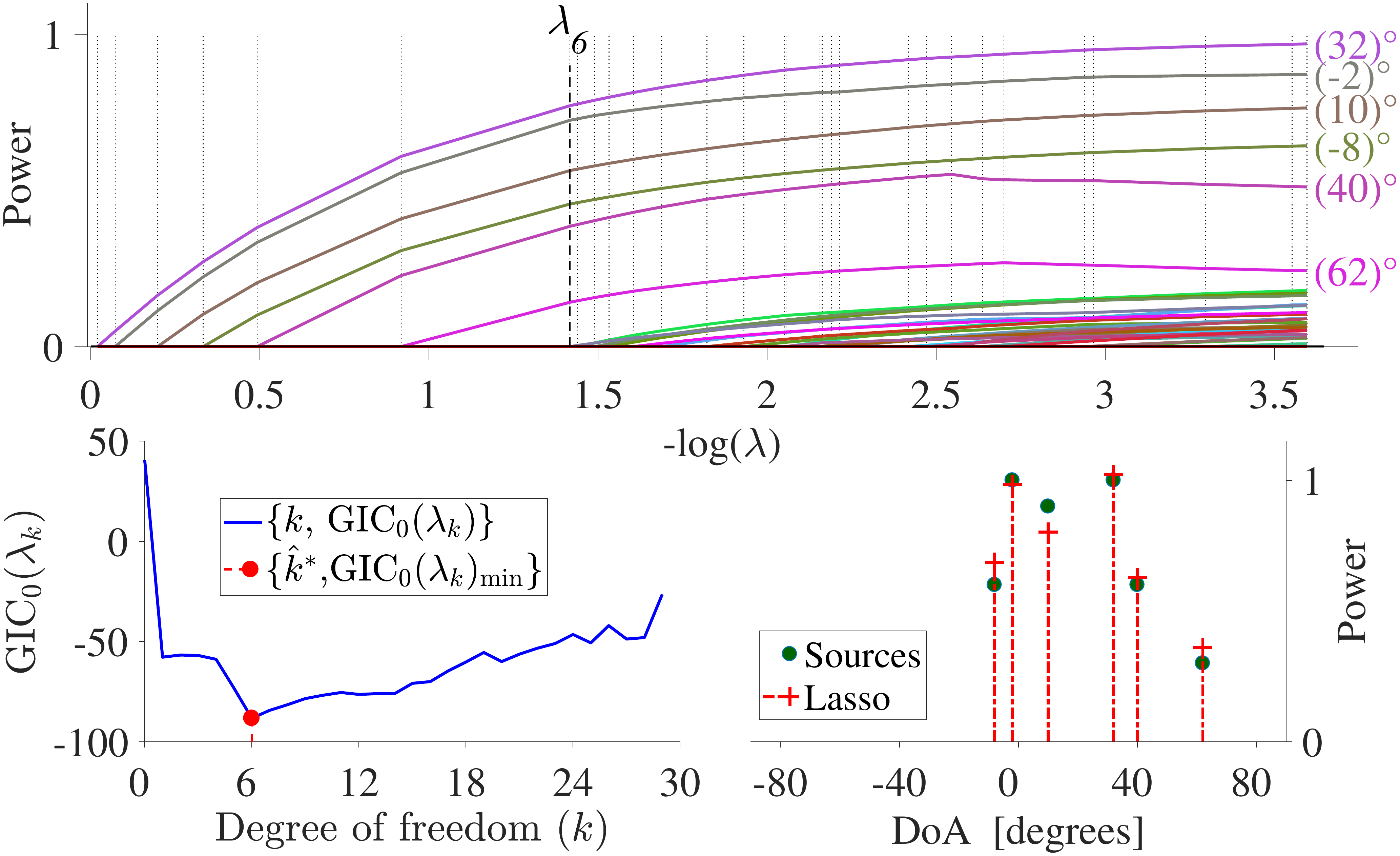}}
	\caption{Upper panel shows the Lasso solution path, i.e., $|[\bebh(\lam)]_j |$, with values of the knots $\lam_k$ as dotted vertical lines, for $k = 0,1,\ldots, K = 29$. Lower left panel shows the GIC scores for $\gam=0$ as a function of $k$. The minimum GIC-score is obtained at $k=6$ giving $\hat k^*=6$ and the magnitudes of the respective  $\hat k^*$-sparse estimate are shown in lower right panel.}
	\label{fig:estKbic}
\end{figure}

For first simulation setup, upper panel in Fig.~\ref{fig:estKbic} depicts the full Lasso solution path with found knots for a snapshot generated from a model, 
where there exists $k^* = 6$ true sources  located at $\bom \theta = (-8,-2,10,32,40,62)^\circ$ with respective powers $(|s_1|,\ldots,|s_6|)= (0.6,1,0.9,1,0.6,0.3)$. In this scenario, we have an SNR level of 15 dB and $\ndim=30$ number of sensors in the ULA.
The lower left panel of Fig.~\ref{fig:estKbic}  depicts the plot of GIC scores  $\mathrm{GIC}_0(\lam_k)$. As can be seen the c-LARS-GIC method chooses a model that has sparsity order $ \hat k^* = k^*=6$. The estimated source powers, $| \bebh_{\hat \setA^*}|$, at DoAs corresponding to found $\hat \setA^*$ are also shown in  lower right panel of Fig.~\ref{fig:estKbic}. Observe that estimated $\hat \setA^*$ coincides with the true $\setA^*$ and that the source powers are estimated with high accuracy. 

\begin{figure}[!t]
	\centering
	\subfloat{\includegraphics[width=0.98\columnwidth]{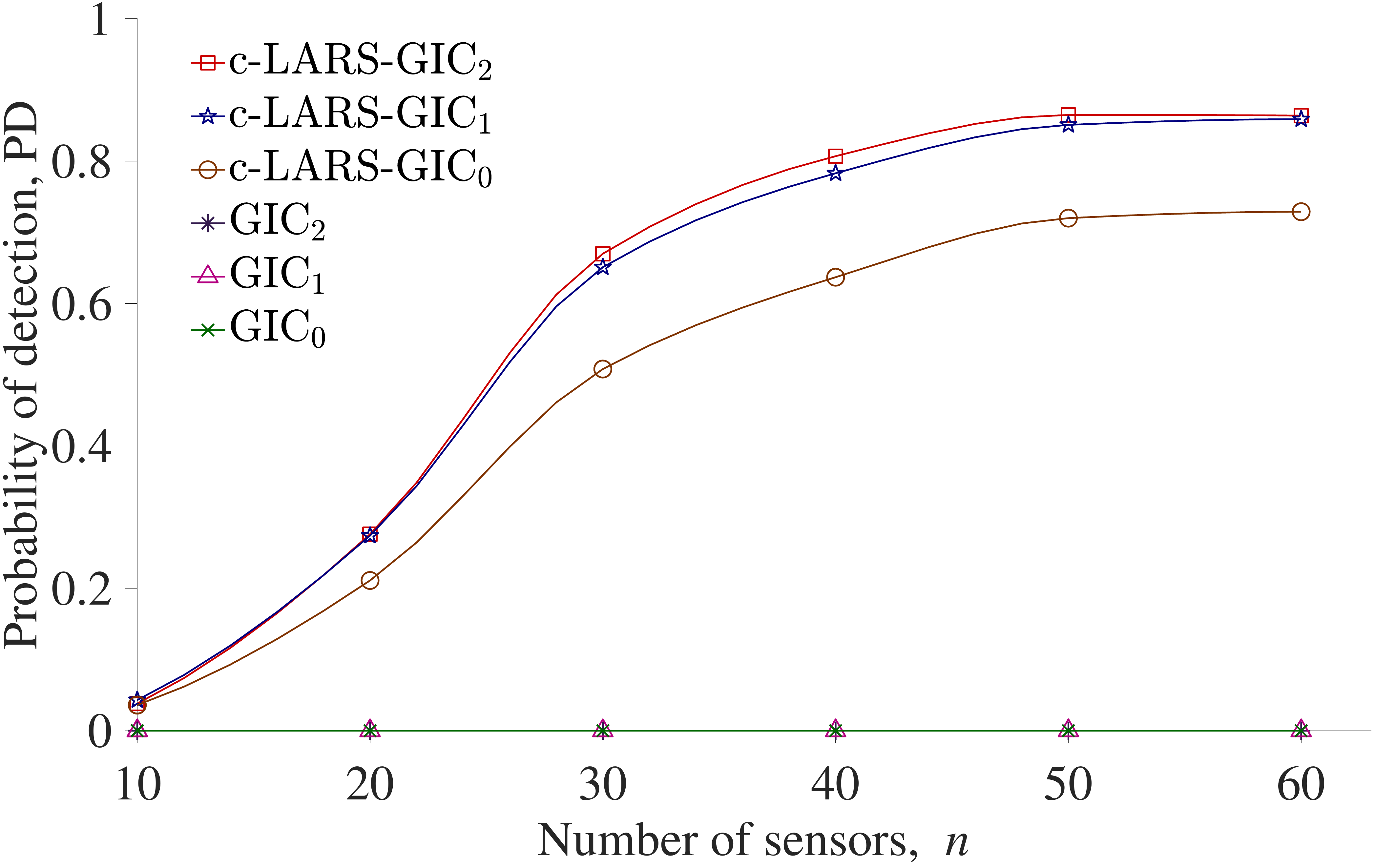}} 
	\caption{Probability of detection  of sparsity order $k^* = 3$  for varying number of sensors $\ndim$. The SNR level is 15 dB.}
	\label{fig:PDvN}
\end{figure}

In  the second simulation we have $k^*=3$ true sources at  $\bom \theta = (-8, 6, 24)^\circ$ with respective powers $(|s_1|,|s_2|,|s_3|) = (0.7, 0.9,  1)$ and the SNR level is 15 dB. When the number of sensors in the ULA is $\ndim=40$, then the c-LARS-GIC$_{\gamma}$ had PD equal to $0.640$, $0.791$, and $0.800$ for $\gamma=0,1,2$, respectively.   
However, conventional grid-based GIC$_\gamma$, as explained in \ref{sec:comp}, failed completely yielding miserable PD $= 0$ in all cases. 		
Fig.~\ref{fig:PDvN} shows the PD as a function of number of sensors $\ndim$ in the ULA.

\begin{figure}[!t]
	\centering
	\subfloat{\includegraphics[width=0.98\columnwidth]{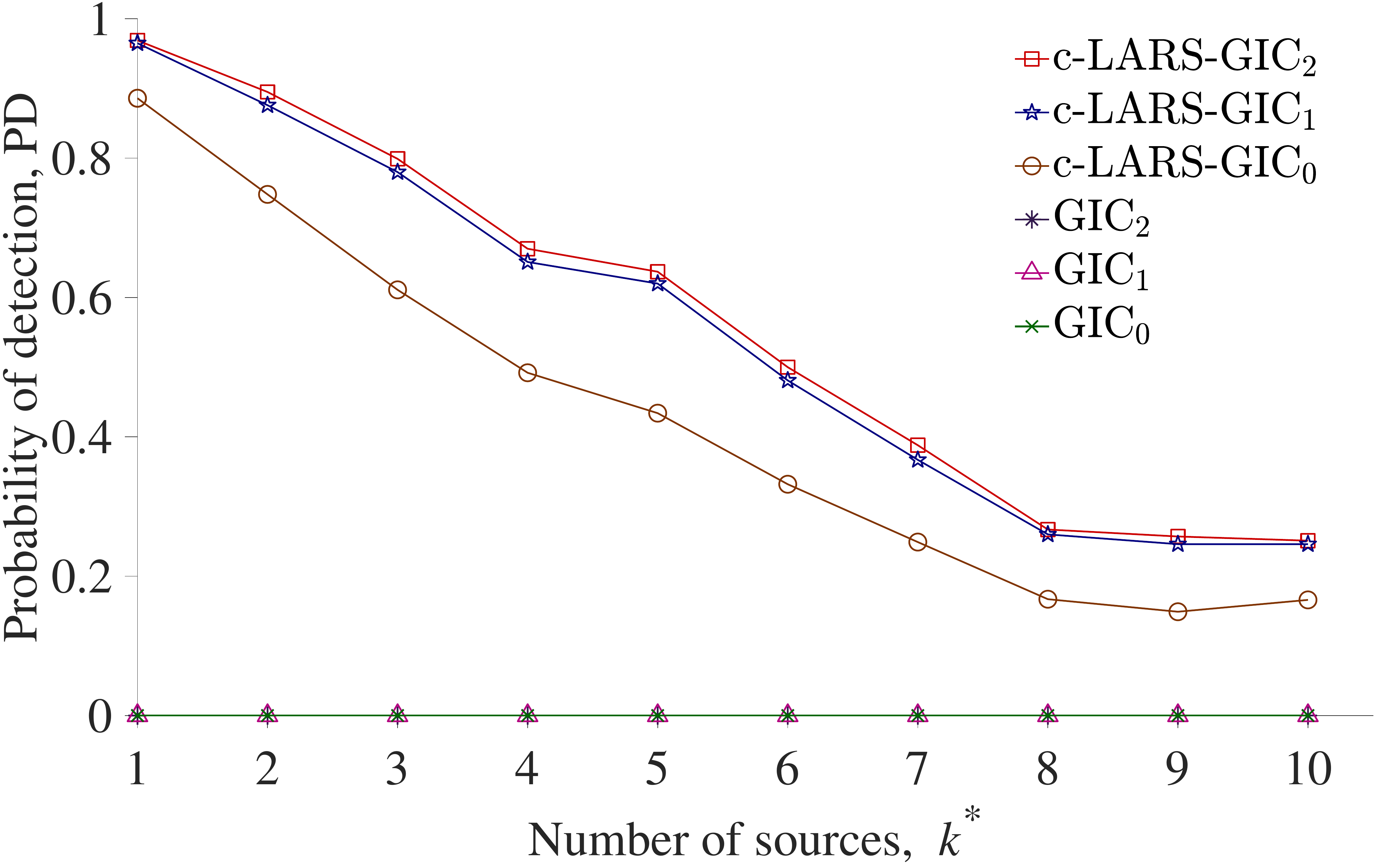}} 
	\caption{Probability of detection (PD)  for varying number of sources $k^*$. The SNR level is 20  dB and $\ndim=40$.}
	\label{fig:probVk}
\end{figure}

\begin{figure}[!t]
	\centering
	\subfloat[]{\includegraphics[width=0.48\columnwidth]{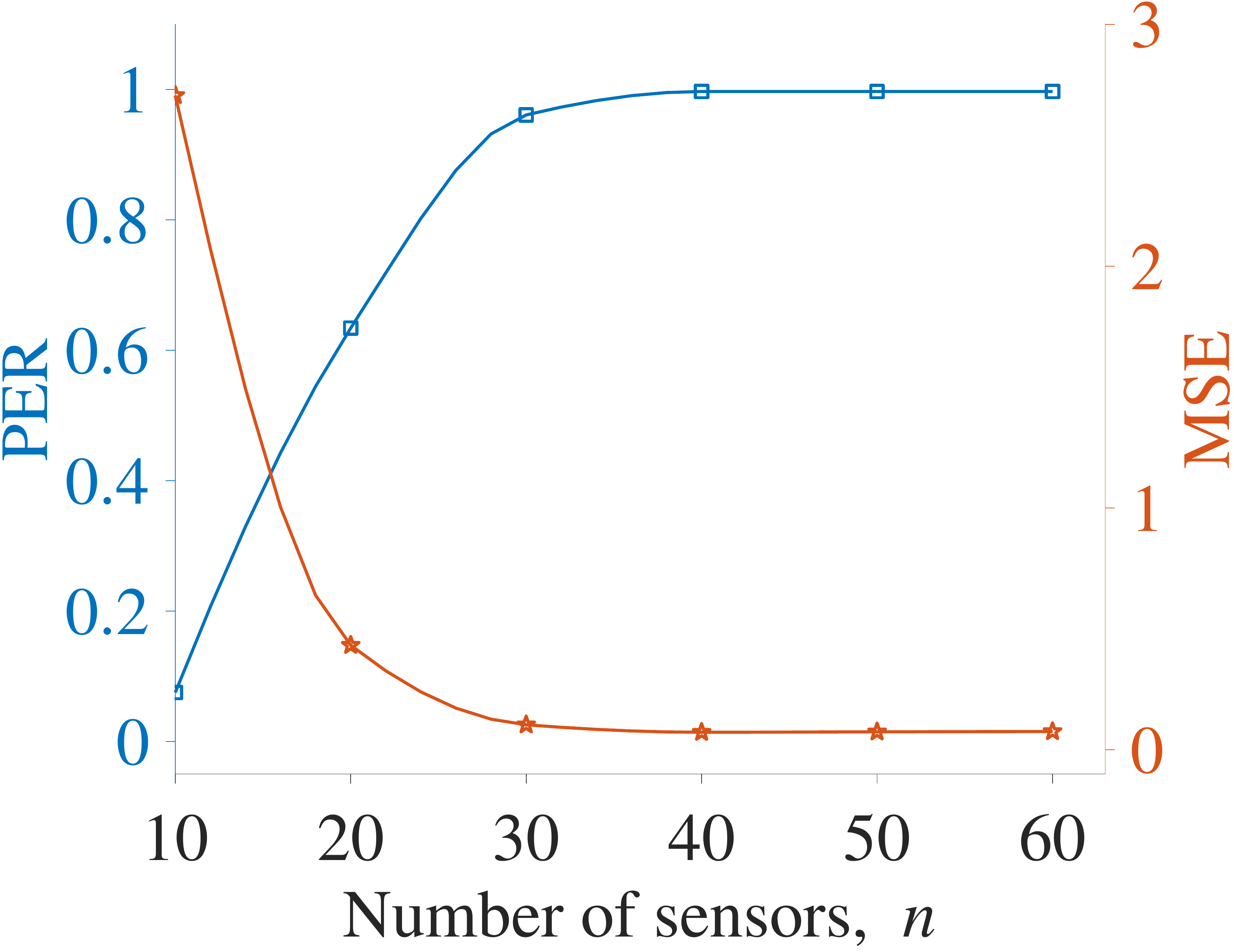}} \quad
	\subfloat[]{\includegraphics[width=0.48\columnwidth]{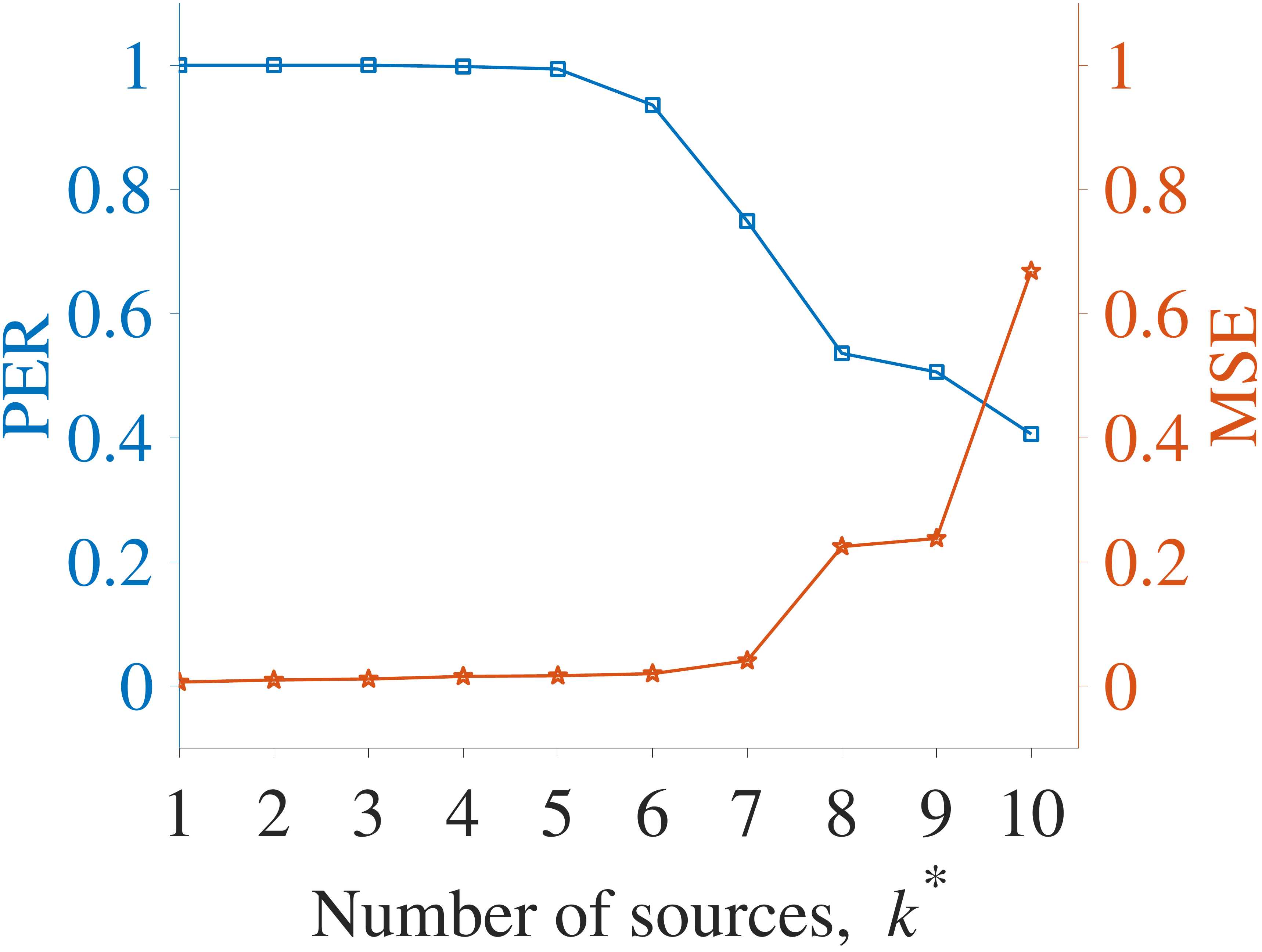}}
	\caption{Probability of exact recovery (PER) of the true support $\setA^*$ and the mean squared error (MSE) for (a) varying number of sensors $\ndim$ and (b) varying number of sources, i.e., sparsity order $k^*$.}
	\label{fig:perVNK}
\end{figure}

Next in the third simulation study,  Fig.~\ref{fig:probVk} shows variation in PD according to the number of  non-zeros (i.e., sources) for fixed $n=40$. An SNR level of 20 dB is used in each scenario when $k^*$ ranges from $k^* = 1, \ldots, 10$.
Herein, a new source at different (straight or oblique) DoA is introduced randomly as the value of $k^*$ increases. We have sources in the following order $\bom \theta = (30,-14,-22,-32,16,-2,56,-30,-8,58)^\circ$ having corresponding magnitudes $(|s_1|, \ldots, | s_{10} |) = (0.8,0.6,0.4,0.6,0.3,0.2,0.9,0.8,0.4,0.9)$.  
It is noticeable that PD decreases as $k^*$ increases. This is expected, since we no longer have $k^* \ll n$ as $n$ is fixed through this simulation.  Moreover, the addition of new sources, especially at oblique angles, results in increased mutual coherence which lowers estimation accuracy \cite{tabassum2018sequential}. These  aspects account for the variation in both probabilities PD and PER of c-LARS-GIC for changing $n$ and sparsity order $k^*$, as visible in both Fig.~\ref{fig:probVk} and Fig.~\ref{fig:perVNK} which display the respective  changes in PER and MSE for second and third simulation setups, respectively.  
Note, however, that the conventional grid-based GIC approach fails to detect the true sparsity level in all the cases. 

\section{Discussions and Conclusions} \label{sec:concl} 

In this paper, we developed the c-LARS based GIC method for finding the sparsity order $k^*$ and corresponding $k^*$-sparse signal estimate. We compared the c-LARS-GIC method to conventional grid-based GIC approach in single-snapshot source localization based on compressed beamforming. 

c-LARS-GIC finds precise values of the penalty parameter where a new variable enters the Lasso path and computes the GIC values for the corresponding active sets. 
The conventional grid-based GIC approach almost always failed as it heavily depends upon the chosen grid. 
Simulation studies illustrated that 
c-LARS-GIC method attains high detection rates 
even at low SNR level of 15 dB. Moreover, the PER and MSE values follow the same pattern and illustrate that c-LARS-GIC is also able to perform accurate sparse signal reconstruction (SSR). 

In conclusion, the results illustrate the potential usage of c-LARS-GIC method in detecting the rank of the signal subspace, i.e., accurate sparsity (model) order selection. Furthermore, its ability to  find the  true support  and magnitudes estimation in SSR task makes it favorable option. Finally, our  software package, containing c-LARS-GIC code along with algorithms of \cite{tabassum2018sequential}, is freely available at \cite{saenlars}.

\section*{Acknowledgment}

The research was partially supported by the Academy of Finland grant no. 298118 which is gratefully acknowledged.

\bibliographystyle{IEEEtran}
\bibliography{eusipco18ref}

\end{document}